\documentclass[aps,prl,reprint,superscriptaddress]{revtex4-1} 
\usepackage{graphicx}
\usepackage{longtable}

\def\lsim{\, \rlap{$<$}{\lower 1.1ex\hbox{$\sim$}}\,}
\begin{document}

\preprint{NITS-PHY-2014004}

\title{Big Bounce Genesis}

\author{Changhong~Li}
\email[]{chellifegood@gmail.com}
\affiliation{Department of Physics, Nanjing University, \\
22 Hankou Road, Nanjing, China 210093}
\author{Robert~H.~Brandenberger}
\email[]{rhb@hep.physics.mcgill.ca}
\affiliation{Ernest Rutherford Physics Building, McGill University, \\
3600 rue University, Montreal, QC H3A 2T8, Canada}
\author{Yeuk-Kwan E. Cheung}
\email[]{cheung.edna@gmail.com}
\affiliation{Department of Physics, Nanjing University, \\22 Hankou Road, Nanjing, China 210093}

\begin{abstract}
We report on the possibility of using  dark matter particle's mass and its interaction cross section as a smoking gun signal of the existence of a Big Bounce at the early stage in the evolution of our currently observed universe.  
A model independent study of dark matter production in the pre-bounce contraction and the post-bounce expansion epochs of the bounce universe reveals 
a new venue  for achieving the  observed relic abundance of our present universe, in which a significantly smaller amount of dark matter with a smaller cross section -- as compared to the prediction of  Standard Cosmology -- is produced and  the information about the bounce universe evolution is preserved by the out-of-thermal-equilibrium process. 
Once the value of dark matter mass and interaction cross section are obtained  by direct detection in laboratories,  this alternative route becomes a signature prediction of the bounce universe scenario.
\end{abstract}


\maketitle


How to extend  the standard cosmology beyond Big Bang Nucleosynthesis (BBN) -- yet compatible with the array of high-precision observations as well as the high energy experimental findings -- is one of the most interesting intellectual challenges.

In standard cosmology inflation was introduced to resolve the {\it flatness problem} and {\it horizon problem} 
in accordance with the homogenous and isotropic conditions of the big bang~\cite{Guth:1980zm}. 
Moreover,  inflation succeeds in generating the scale-invariant power spectrum of primordial perturbations 
 which agrees well with the current array of observations~\cite{Komatsu:2010fb, Planck:2013kta, Ade:2014xna}.  
Being an effective scalar field theory, inflationary models inevitably inherits  the Big Bang Singularity~\cite{Borde:1993xh}.

Since this realization  there has been a concordance of effort to address the issue of Big Bang Singularity%
--notably the String Gas Cosmology~\cite{Brandenberger:1988aj}, 
the Pre-Big-Bang Scenario~\cite{Gasperini:1992em, Gasperini:2002bn}, and matter-dominated contraction preceding the 
Big Bang~\cite{wands1998}\footnote{And many variations on the theme~\cite{Finelli:2001sr, Creminelli:2006xe,Cai:2007qw,%
Cai:2008qw,Wands:2008tv,Brandenberger:2009yt,Lin:2010pf,
Avelino:2012ue,Bhattacharya:2013ut}.}, the Brane-World scenario~\cite{Dvali:1998pa} and the Ekpyrotic Universe~\cite{Khoury:2001wf}. 
The common theme of these attempts was to embed inflationary paradigm into a more fundamental theory of quantum gravity, e.g. string theory and loop quantum gravity, resorting to more complex duality and/or  symmetry properties of the mother theories to resolve the big bang singularity. 

Recently a {\em{stable}} as well as  scale-invariant power spectrum of primordial density  perturbations 
is finally  obtained~\cite{Li:2011nj, Li:2013bha} in the bounce universe scenario. 
The ``Bounce Cosmology'' postulates that there exist a phase of 
matter-dominated contraction before the big bang~\cite{wands1998} in which the matter content of the universe comes into thermal contact--resulting in a scale invariant spectrum--before a subsequent expansion after the ``big bounce.''
In view of this development  we are  motivated to work out 
further experimental or observational predictions from the 
bounce universe scenario.

\begin{figure}[htp!]
\centering
\includegraphics[width=0.48\textwidth]{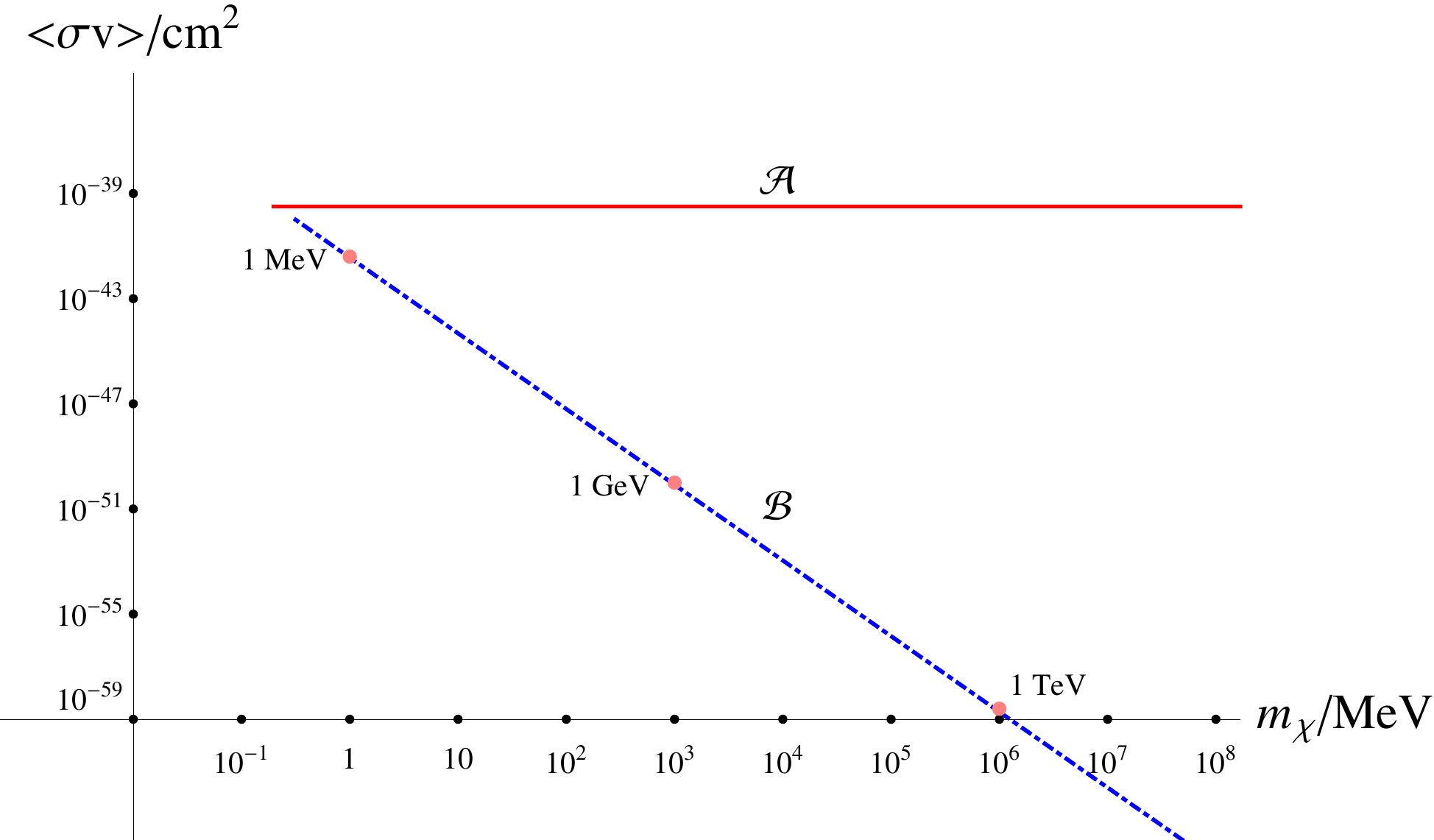}
\caption{Cosmological constraints on $\langle \sigma v\rangle$ and $m_\chi$ in the Bounce Universe Scenario. For comparison the prediction from standard cosmology~\cite{Kolb:1990vq, Dodelson:2003ft} is displayed on
branch $\mathcal{A}$. 
A few dark matter particle masses of common interest and their predicted values of $\langle \sigma v \rangle$ in Big Bounce Scenario are  marked.
}
\label{fig:cbplog}
\end{figure}

In this Letter we  report a signature prediction of bounce scenario by studying the out-of-equilibrium dark matter production  in the pre-bounce contraction and the post-bounce expansion, and subsequent freeze-out process in the post-bounce radiation-dominated epoch. 
A characteristic relation, depicted by a blue line marked 
``\,$\mathcal{B}$\,'' in Fig.~\ref{fig:cbplog}, 
between the dark matter mass and its interaction cross section 
is obtained, which is significantly different from the analogous relation obtained from the standard cosmology, marked ``\,$\mathcal{A}$\,'' in the same figure.~\footnote{Note that even in the standard cosmology, the non-thermal production of the dark matter could also happen, which has been utilized in~\cite{Chung:1998ua} to test non-standard cosmology.   The relation among dark matter mass and cross section predicted in the standard cosmology  is, however,  generically different from the predictions from the bounce universe scenario.    
In order  to highlight  the new relation  we found 
between dark matter mass and cross section in the bounce universe scenario from  non-thermal production, we restrict our comparison between the non-thermal production  and the thermal production. In the latter case of thermal production the relation is the same as in the standard cosmology.}
The characteristic relation of mass and cross section  can  then  be checked against the data from worldwide effort of direct detections, 
LUX~\cite{Akerib:2013tjd}, XENON10~\cite{XENON08}, XENON100~\cite{Aprile:2012nq,Xe100spin13}, CoGENT~\cite{CoGeNT11}, DAMA~\cite{DAMA1,DAMA10, DamaSasso13}, PICASSO~\cite{PICASSO09,PICASSO11}, PANDAX~\cite{Xiao:2014xyn} as well as AMS~\cite{Battiston:2014pqa,Aguilar:2014mma}, ATIC~\cite{ATIC}, PAMELA~\cite{Adriani:2013uda} and EGRET~\cite{EGRET}, 
to determine whether or not our universe obeyed the evolution prescribed by the standard cosmology or by the bounce cosmology~\cite{Cheung:2014pea}. 

Encouraged  by our findings, a few other research 
teams~\cite{%
Li:2014msi,%
Quintin:2014oea,
Wan:2014fra,
Cai:2014bea,
Liu:2014tda,
Li:2014qwa,
Cai:2014hja,
Cai:2014xxa,
Hu:2014aua, 
Li:2014cka,
Xia:2014tda,
Cai:2014zga
} 
have worked out various predictions from the bounce universe, 
most notably a paper by  Xia et al~\cite{Xia:2014tda} recently published  in PRL and  featured in news APS focus~\cite{Lindley:2014gia}, 
in which the tensor-scalar ratio of a particular bounce universe model is worked out and fit to WMAP and BISEP2 data. 

Our prediction is of particle physics nature, compared with the other attempts mentioned above  of using cosmological data to test the bounce scenario,  our predictions enjoy the high precision and repeatability of high energy experiments,  as well as independence 
in methodologies between making predictions and testing predictions. 
Our predictions have no free parameters: for every dark matter mass we predict one unique interaction cross section. This is to be contrasted with standard model cosmology in which a wide range of mass can satisfy the relic abundance provided the cross section is of the weak interaction strength.  
The mass of dark matter particle can be obtained in the many dark  matter experiments and at LHC, if the cross section--an experimental precision within an order of magnitude suffices to place it significantly far away from the standard model prediction--turns out not to be what we predict in this Letter the bounce universe scenario can be casted into serious doubt.  

\paragraph{A signature prediction from the bounce universe:}

\begin{figure}[htp!]
\centering
\includegraphics[width=0.48\textwidth]{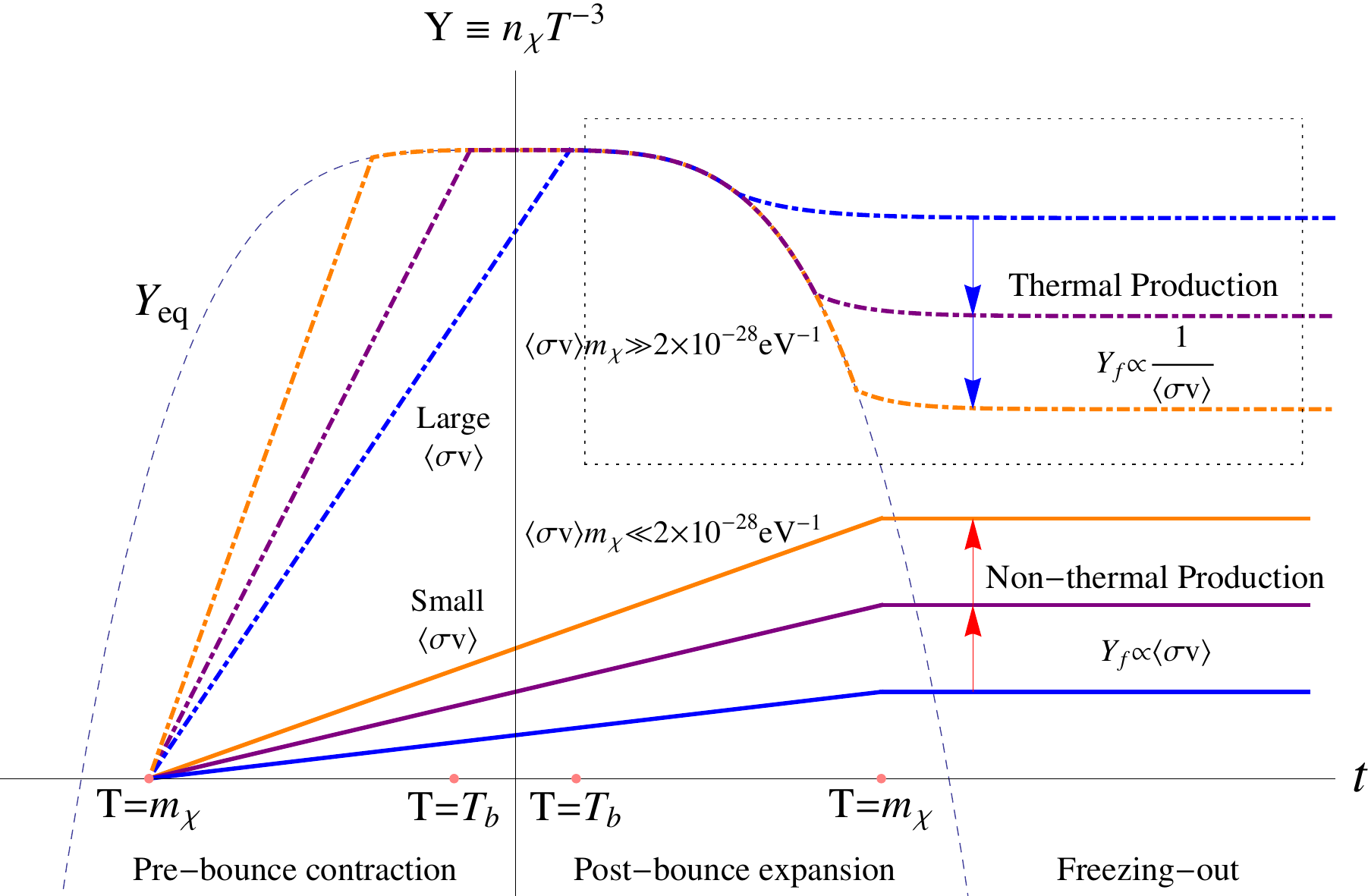}
\caption{A schematic plot of the time evolution of dark matter
in a generic bounce universe scenario. Two pathways of producing dark matter yet satisfying current observations -- thermal
production (which is indistinguishable from standard cosmology) and non-thermal production (characteristic to bounce universe) -- are illustrated. The horizontal axis indicates both the time, $t$, as well as the temperature, $T$,  of the cosmological background.}
\label{fig:Relics}
\end{figure}

By investigating the production process  of dark matter in  
the pre-bounce contraction and the post-bounce expansion 
epochs of a generic bounce universe, we find that, in the big bounce scenario, dark matter production can be extended  beyond the Big Bang, as shown in Fig.~\ref{fig:Relics}. 
Furthermore an out-of-thermal-equilibrium production of dark matter is allowed which encodes information of early universe evolution, marked the ``non-thermal production'' in Fig.~\ref{fig:Relics}. 
Specifically
\begin{itemize}
\item it predicts a characteristic relation governing a dark matter mass and interaction cross section, depicted by a blue line marked 
``\,$\mathcal{B}$\,'',  in Fig.~\ref{fig:cbplog};
\item a factor of $1/2$ in thermally averaged cross section, as compared to nonthermal production in standard cosmology, in needed for creating enough dark matter particle to satisfy the currently observed relic abundance because  dark matter is being created during the pre-bounce contraction, in addition to the post-bounce expansion.
\end{itemize}

\paragraph{A model independent approach to dark matter study in bounce universe:}
As shown in Fig.~\ref{fig:Relics}, we divide the bounce 
(See~\cite{Novello:2008ra, Brandenberger:2012zb} 
for recent reviews.) schematically into three stages to 
facilitate a model independent analysis: 
\begin{itemize}
\item
Phase I: the pre-bounce contraction, in which $H<0$ and $m_\chi< T < T_b$;
\item
{
Phase II: the post-bounce expansion, 
in which $H>0$ and $m_\chi< T < T_b$;
} 
\item
{
Phase III:  the freeze-out of dark matter,  in which  $H>0$ and  $m_\chi \, >\,  T$;
}
\end{itemize}
where $m_\chi$   is  the mass of DM, $\chi$, and $H$ the Hubble parameter  taking positive value in expansion and negative value in contraction. 
$T$ and $T_b$ are the temperatures  of the cosmological 
background and of the bounce point, respectively. 
The bounce  point,  connecting Phases I and II,  
is highly model-dependent.  
And  its detailed modeling is an actively researched subject but luckily it  is  sub-leading effect to our analysis of dark matter production as long as its time scale is short and the bounce is smooth as it is usually assumed. 
Therefore, given that the entropy of universe is conserved around the bounce point~\cite{Cai:2011ci}, the relic abundance of $\chi$, ${n_\chi}$, at the end of the pre-bounce contraction (denoted by $-$)  is equal to the initial abundance of the post-bounce expansion (denoted by $+$),
\begin{equation} \label{eq:match}
{n_\chi}^{(-)} (T_b) = {n_\chi}^{(+)} (T_b)~. 
\end{equation} 
Without loss of generality  the  number density of dark matter particles, $n_\chi$,  is set to zero at the onset of the pre-bounce contraction phase~\footnote{
Even if $n_\chi$ is not set to zero it would only raise the characteristic curve by a constant background but does not change the shape of the curve.
}
: 
\begin{equation}     \label{eq:ini}
{n_\chi}^{(init)} =0. 
\end{equation} 

The  evolution of dark matter in the bounce  scenario can be  analyzed using  the Boltzmann equation, 
\begin{equation} \label{eq:nsf}
\frac{d(n_\chi a^3)}{a^3dt}=\bar{\langle\sigma v\rangle}\left[\left(n_\chi^{(0)}\right)^2-n_\chi^2\right]~, 
\end{equation}
where $n_\chi^{(0)}$ is the equilibrium number density of dark matter, 
$a$,  the scale factor of the cosmological background, 
and, $\bar{\langle\sigma v\rangle}$, the thermally averaged cross section with temperature dependence. 

We model the interactions of the dark matter with a light boson by 
$\mathcal{L}_{int} = \lambda \phi^{2} \chi^{2}$. 
In the limit $m_\phi\rightarrow 0$ one gets~\cite{Peskin:1995ev},
\begin{equation} \label{eq:sigmavT}
\bar{\langle\sigma v\rangle}=\left\{  
\begin{array} {l}
 {\displaystyle \frac{x^2}{4}\cdot\langle\sigma v\rangle ~, \qquad m_\chi\ll T}  \\ 
 \\ 
  {\displaystyle  \langle\sigma v\rangle ~, \qquad\qquad m_\chi\gg T}    \\
\end{array}     
\right. ,
\end{equation}
where $\langle\sigma v\rangle\equiv \frac{1}{32\pi} \frac{\lambda^2}{m_\chi^2}$ 
and $\bar{\langle\sigma v\rangle}$ is computed  in low temperature limit, $T\ll m_\chi$.

We take the universe in the pre-bounce contraction 
and post-bounce expansion to be radiation dominated, 
$\rho\propto a^{-4}$,  the Boltzmann equation Eq.(~\ref{eq:nsf})
governing the $\phi-\chi$ interaction in the pre-bounce contraction, $m_\chi<T<T_b$, is
\begin{equation} \label{eq:boltzmannI}
\frac{dY_-}{dx}=-f \langle\sigma v\rangle m_\chi(1-\pi^4Y_-^2)~,
\end{equation}
where $f$ is a constant during the radiation-dominated era with  
$f\equiv \frac{m_\chi^2}{4\pi^2} (|H|x^2)^{-1}=1.5\times 10^{26}~eV$, 
being constrained by observations. 
Consequently the Boltzmann equation Eq.(\ref{eq:nsf}) in the post-bounce expansion phase simplifies (with the usual shorthand notations, $Y\equiv\frac{n_\chi}{T^3}$, and $x\equiv\frac{m_\chi}{T}$),
\begin{equation} \label{eq:dYdxII}
\frac{dY_+}{dx}=f\langle\sigma v\rangle m_\chi(1-\pi^4Y_+^2)~. 
\end{equation}
Solving the Boltzmann equations above with the initial condition Eq.(\ref{eq:ini}) and the matching condition Eq.(\ref{eq:match}),
we obtain the complete solution of the dark matter abundance in the post-bounce expansion, 
\begin{equation} 
\label{eq:Yplus}
Y_+=\frac{1-e^{-2\pi^2f\langle\sigma v\rangle m_\chi(1+x-x_b)}}{\pi^2\left(1+e^{-2\pi^2f\langle\sigma v\rangle m_\chi(1+x-2x_b)}\right)}~. 
\end{equation} 

At the end of dark matter production, $T\sim m_\chi$, 
this complete solution can be analyzed in two limits, the large cross section case and the small cross section case: 
\begin{equation}  
\label{eq:Yplus2}
Y_+|_{x=1}=
\left\{  
  \begin{array} {lr}
 {\displaystyle \pi^{-2}, \qquad\qquad~  4\pi^2f\langle\sigma v\rangle m_\chi\gg 1} 
 \\ 
  {\displaystyle 2f\langle\sigma v\rangle m_\chi~, \quad~ 4\pi^2f\langle\sigma v\rangle m_\chi \ll 1}   
  \\
\end{array}     
\right. .
\end{equation}
These two distinct cases point to two possible  paths of  producing  dark matter in the Big Bounce: 
In the case of  large cross section, dark matter is produced in plenty abundance in thermal equilibrium; whereas in the latter case of small cross section, the production is mostly oneway and thermal equilibrium cannot be established. 
We shall call in the rest of the Letter  the former  {\it ``thermal production''} and the latter {\it ``non-thermal production''}, as depicted in Fig.~\ref{fig:Relics}.

With {\it thermal production}, dark matter is  in  thermal equilibrium with the background plasma and  the dark matter abundance tracks the equilibrium until freeze-out, and therefore its  relic density is inversely proportional to the cross section, $\Omega_\chi\propto \langle \sigma v\rangle^{-1}$, see~\cite{Kolb:1990vq, Gondolo:1990dk, Dodelson:2003ft, Scherrer:1985zt, Feng:2008ya} for the similar pathway in the standard cosmology. And all information about the early universe evolution is washed out. 

In the {\it non-thermal} path dark matter is produced completely out of thermal equilibrium and its relic density is proportional to the cross section, $\Omega_\chi\propto \langle \sigma v\rangle$.
It is remarkable that this novel venue is able to encode the information about  the cosmic evolution of the bounce universe, 
in big contrast to standard cosmology. And from this information we extract our testable prediction of the bounce universe~\footnote{A similar mechanism but within standard cosmology's reheating framework has been analyzed in~\cite{Chung:1998ua, Chung:1998rq, Drees:2006vh}.}.

\paragraph{Freeze-out:}
To determine the final abundance of dark matter observed 
today we  study the details of the freeze-out processes  
in each of the  two cases. 

As the temperature of  universe continues to fall in the radiation dominated era with $m_\chi < T$,  dark matter undergoes thermal decoupling while the cosmos  is still in expansion (Phase III in Fig.~\ref{fig:Relics}). 
The Boltzmann equation Eq.(\ref{eq:nsf}) becomes
\begin{equation}  \label{eq:dYdx}
\frac{d Y}{d x}=4f\langle\sigma v\rangle m_\chi\left(\frac{\pi}{8}xe^{-2x}-\pi^4\frac{Y^2}{x^2}\right)~,
\end{equation}
where the first term on the right hand side of Eq.(\ref{eq:dYdx}) is subdominant and hence discarded for $x>1$. 
Integrating it from $x=1$ to $x\rightarrow\infty$, we obtain the relic abundance of dark matter after freeze-out:
\begin{equation}  \label{eq:dYdxsol}
Y_f\equiv Y|_{x\rightarrow\infty}=\frac{1}{4\pi^4f\langle\sigma v\rangle m_\chi +(Y_+|_{x=1})^{-1}}~.
\end{equation}

Again, as can be seen from Eq.(\ref{eq:dYdxsol}), there are two distinctive outcomes of the freeze-out process.

\paragraph{Strong freeze-out:}  If the initial abundance of dark matter at the onset of the freeze-out process is very large, $Y_+|_{x=1}\gg (4\pi^4f\langle\sigma v\rangle m_\chi)^{-1}$, the relic abundance of dark matter becomes independent of the initial abundance, and is inversely proportional  to the cross section after freeze-out,  
\begin{equation}
Y_f= \frac{1.71 \times 10^{-29} eV^{-1}}{\langle\sigma v\rangle m_\chi}~.
\end{equation}

As the relic abundance tracks the equilibrium until freeze-out, all information of the early universe is washed out. This is analogous to the standard cosmology in which the {\it strong} freeze-out condition is always 
assumed~\cite{Kolb:1990vq, Dodelson:2003ft}: the well-known WIMP miracle~\cite{Scherrer:1985zt}, WIMP-less miracle~\cite{Feng:2008ya} all share this assumption. 

\paragraph{Weak freeze-out:} If, on the other hand, $Y_+|_{x=1}\ll(4\pi^4f\langle\sigma v\rangle m_\chi)^{-1}$ and the density of dark matter is too low  to pair-annihilate during the thermal decoupling. The relic abundance of dark matter after freeze-out in this limit is just the initial abundance at the onset of the freeze-out process,  
\begin{equation}
Y_f= Y_+|_{x=1}~.
\end{equation}

In this way the final relic abundance of dark matter observed today encodes the information of the early universe dynamics, if dark matter is produced out of thermal equilibrium and then undergoes a weak freeze out when it decouples from the primordial plasma. This is reason for the existence of  the signature prediction of the bounce universe. 

\paragraph{Summary:} 
Two different production routes 
and two distinct 
freeze-out outcomes give four combined possibilities but resulted in only two logically viable predictions, summarized in
Table~\ref{tab:dmfreezeout}~\footnote{Two other combined possibilities are absent in Table~\ref{tab:dmfreezeout} since the production conditions coincide with the freeze-out conditions as $\lambda$ is temperature independent. However, if the cross-section is assumed to be temperature independent~\cite{CKL_wip},  there exists another possibility with thermal production and  weak freeze-out.}%
.
\begin{table}[ht!]  
\vspace{-0.5cm}
\caption{\label{tab:dmfreezeout}Relic abundance of DM after strong/weak freeze-out processes.}
\begin{center}
\begin{tabular}{|c|c|c|}
\hline
& thermal ${(\mathcal{A})}$ & non-thermal ${(\mathcal{B})}$ \\
\hline
strong  & $Y_f=\displaystyle{\frac{
1.71\times 10^{-29} eV^{-1}}{\langle\sigma v\rangle m_\chi}}$& --- \\
\hline
weak  & --- &
$Y_f=3\times10^{26} eV\, \langle\sigma v\rangle m_\chi $\\
\hline
\end{tabular}
\end{center}
\vspace{-0.5cm}
\end{table}

\paragraph{The signature of the Big Bounce:}
The novel feature of dark matter production in the bounce universe scenario arises from  the branch~$\mathcal{B}$. Specifically
\begin{itemize}
\item it predicts a characteristic relation governing a dark matter mass and interaction cross section, depicted by a blue line marked 
``\,$\mathcal{B}$\,'',  in Fig.~\ref{fig:cbplog};
\item a factor of $1/2$ in thermally averaged cross section as compared to standard cosmology's nonthermal production cross section is needed for creating enough dark matter particle to satisfy the currently observed relic abundance, because dark matter is created during the pre-bounce contraction, in addition to the post-bounce expansion.
\end{itemize}
 
Due to the smallness of the interaction  cross-section, dark matter particles are produced {\it non-thermally}, and its relic abundance never reaches  thermal equilibrium but retains the  information of cosmic dynamics  of the bounce universe. 

Therefore,  any information relating  dark matter's  mass and interaction cross section extracted from this path would enable us to confirm  the existence of a bounce process,  or the lack of it, in our early universe. 

\paragraph{Cosmological constraints:} 
We will now constrain the dark matter's mass and interaction cross section by imposing the current  value of $\Omega_\chi$~\cite{Dodelson:2003ft}, $\Omega_\chi=1.18\times 10^{-2}eV\times m_\chi Y_f  =0.26$. 
The cosmological constraint  on ${\langle\sigma v\rangle}$ and  $m_\chi$ for the two branches are summarized in Table~\ref{table:results} and plotted in Fig.~\ref{fig:cbplog}.
\begin{table}[htdp]  
\vspace{-0.5cm}
\caption{\label{table:results} Cosmological constraints on ${\langle\sigma v\rangle}$ and 
 $m_\chi$ in the Bounce Universe Scenario.}
 \vspace{-0.0cm}
\begin{center}
\begin{tabular}{|c|c|}
\hline 
   &    $\Omega_\chi=0.26$ at present \\ 
\hline
Branch $\mathcal{A}$~:~ &   
${\langle\sigma v\rangle}=0.31\times 10^{-39}cm^2$~, $m_\chi\gg 216~eV$\\ 
\hline
Branch ${\mathcal{B}}$~:~ &  
${\langle\sigma v\rangle}=7.2\times 10^{-26}m_\chi^{-2} $~,~  $m_\chi\gg 432~eV$\\
\hline
\end{tabular}
\end{center}
\vspace{-0.5cm}
\end{table}

The concrete relation, 
${\langle\sigma v\rangle}\propto m_\chi^{-2}$, 
depicted in  branch~$\mathcal{B}$, can be tested in the forthcoming array of dark matter direct detection experiments 
with improved precision. 
Even though the cross section predicted by the 
branch~$\mathcal{B}$ for dark matter with large masses, 
say $m_\chi > 1~GeV$, is smaller than the current detectable limit,   it is still possible to be detected in the near future with rapidly improved detection technologies and exponentially increasing target sizes~\cite{Aprile:2012nq, Akerib:2013tjd}. 

If the experimentally measured values of $m_\chi$ and 
$\langle {\sigma v} \rangle$ fall on this curve then 
it is a strong indication that the our universe undergoes a Big Bounce--instead of the inflationary scenario as postulated in standard cosmology! 
Accordingly, with the definition of $\langle\sigma v\rangle$, 
the branch~$\mathcal{B}$ predicts a fixed value of coupling constant, $\lambda=0.27\times 10^{-11}$ for satisfying 
the currently observed value of $\Omega_\chi$\footnote{Because $\lambda$ is independent of the value of $m_\chi$ in 
the branch~$\mathcal{B}$, the upper limit of $m_{\chi}$ could be extended to a larger value dramatically. This is to be contrasted with WIMP and other scenarios in standard cosmology in which  the upper limit of $m_\chi$  is constrained by the unitary limit of the coupling constant~\cite{Griest:1989wd}}.

Before closing this section one remark concerning the dependence of our analysis on the energy scale of the Big Bounce, $T_{b}$,  is warranted. 
We have taken $m_\chi<T_b$  the energy scale of the bounce
 appears only in subheading order in the relic 
 abundance Eq.(\ref{eq:Yplus}). 
This, in turn, implies that the prediction of the 
branch~$\mathcal{B}$ is robust  against any variations in the energy scale of the Big Bounce. 
We also notice that  if  the Big Bounce is cold,  
 $T_b \ll m_\chi$, the bounce scale appears in the leading order of the relic abundance, $Y_f\propto e^{-2x_b}(1+x_b)$ and
 $x_b\gg 1$.  This fact  can then  be used to constrain the 
 energy scale of  the Big Bounce~\cite{CKL_wip}.

\paragraph{Conclusion and Outlook:}
We report in this Letter a possibility of experimentally testing the recently proposed Big Bounce Scenario using dark matter direct detections. 
Dark matter can be produced out of thermal equilibrium  during 
the pre-bounce contraction and post-bounce expansion, 
the cross section is comparably smaller than WIMP and hence 
calls for larger dark matter mass in order to satisfy the current dark matter density constraint. 
A characteristic relation of $m_{\chi}$ and 
$\langle\sigma v \rangle$ obtained (the blue solid line in Fig.~\ref{fig:cbplog})
can be turned into a  discriminating signal for the existence, or the lack thereof, of the Big Bounce at an early stage of cosmic evolution.  

If, however, the experimentally determined values of dark matter mass and interaction cross section do not fall onto  this theoretical curve then the bounce universe scenario has no reason to be  a  preferred scenario  over the standard inflationary cosmology. 
 Even though one can consider some modifications of our model independent analysis by including detailed modeling of the bounce process, the overall shape of the 
$m_{\chi}$--$\langle\sigma v \rangle$ 
curve will not be changed qualitatively. 

As we discover a whole new venue for dark matter production 
in the bounce scenario because of the existence of the phase of pre-bounce contraction, ``reheating'' is no longer called for in the bounce universe. 
It is therefore natural to expect that  baryon  genesis
can similarly enjoy a major rethinking in the bounce universe~\cite{CKLR-M_wip}.

\begin{acknowledgments}
We would like to thank Jin U Kang,  Kai-Xuan Ni, and Konstantin Savvidy for many useful discussions.
This research project has been supported in parts by the Jiangsu Ministry of Science and Technology under contract BK20131264. 
We also acknowledge 985 Grants from the Ministry of Education, and the Priority Academic Program Development for Jiangsu Higher Education Institutions (PAPD).

\end{acknowledgments}

%

\end{document}